\documentclass[review]{elsarticle}

\usepackage{hyperref}

\journal{Journal of \LaTeX\ Templates}

\usepackage{graphicx}
\usepackage[utf8]{inputenc}
\usepackage{natbib}
\usepackage{amsfonts}
\usepackage{amssymb}
\usepackage{mathrsfs}
\usepackage{amsmath}
\usepackage{mathtools}
\usepackage{tikz}
\usepackage{subcaption}
\usepackage{pdflscape}
\usepackage{afterpage}
\begin{document}

\begin{frontmatter}

\title{The Effects of Air Quality on the Spread of the COVID-19 Pandemic in Italy: An  Artificial Intelligence Approach.}

\author{Andrea Loreggia}
\address{European University Institute}

\author{Anna Passarelli}
\address{University of Padova}

\author{Maria Silvia Pini}
\address{University of Padova}

\begin{abstract}
The COVID-19 pandemic considerably affects public health systems around the world. The lack of knowledge about the virus, the extension of this phenomenon, and the speed of the evolution of the infection are all factors that highlight the necessity of employing new approaches to study these events. Artificial intelligence techniques may be useful in analyzing data related to areas affected by the virus.
The aim of this work is to investigate possible relationships between air quality and the spread of the pandemic. We also evaluate the performance of machine learning techniques on predicting new cases. Specifically, we describe a cross-correlation analysis on daily COVID-19 cases and environmental factors, such as temperature, relative humidity, and atmospheric pollutants. 
Our analysis confirms a significant association of some environmental parameters with the spread of the virus. This suggests that machine learning models trained using environmental parameters might provide accurate predictions about the number of infected cases. Predictive models may be useful for helping institutions in making decisions for protecting the population and contrasting the pandemic. Our empirical evaluation shows that temperature and ozone are negatively correlated with confirmed cases (therefore, the higher the values of these parameters, the lower the number of infected cases), whereas atmospheric particulate matter and nitrogen dioxide are positively correlated. We developed and compared three different predictive models to test whether these technologies can be useful to estimate the evolution of the pandemic.  
\end{abstract}


\end{frontmatter}


\section{Introduction}
The new coronavirus SARS-CoV-2 is responsible for the respiratory disease named COVID-19. It was first identified on the $9^{th}$ of January 2020 by the Municipal Health Commission of Wuhan (China) which reported to the World Health Organization (WHO) a cluster of pneumonia cases of unknown origin in the city of Wuhan, in the Chinese province of Hubei. The spread of COVID-19 was then declared a global pandemic by WHO on  the $11^{th}$ of March 2020 \cite{world2020director}. On the $8^{th}$ of April 2021, the World Health Organization reported the number of confirmed global cases of COVID-19 exceeds 132 million, with more than 2.8 million deaths. At that date in Italy more than 3.6 million positive cases and almost 112 thousand deaths have been recorded \footnote{https://covid19.who.int/}.

The scale of the public health emergency caused by COVID-19 has no precedent in recent decades and it will surely have serious social and economic consequences. Indeed, the rapid spread of this global pandemic has immediately raised urgent issues, which need a coordinated study to slow down the evolution of the disease. In this context, Artificial Intelligence (AI) techniques can represent a great support for government institutions and health organizations in order to provide information on the mechanisms which describe how the virus spread and, possibly, on the methodologies to be adopted to contrast it in the most effective way. If properly implemented and used, machine learning algorithms can help in analyzing data relating to some areas affected by the infection (see for instance, \cite{vaishya2020artificial,jamshidi2020artificial,lalmuanawma2020applications}). Analyzing the available historical data,  machine learning models can be trained to predict possible developments of the pandemic as well as the  impacts on the population.

Understanding a complex system such as the spread of the pandemic is an important challenge in a country's sustainable development process. The concept of sustainable development has spread widely in recent decades and generally consists of a combination of three goals: the social goal, the economic goal, and the environmental goal. Policy makers play a key role in these scenarios in order to reach these goals, but they have to be properly informed in order to make the right decision. This process may be helped and improved by adopting the right technology such as artificial intelligence techniques.
During the last months, the attention of many researchers has moved to this new challenge that has involved the entire planet. For instance, a group of researchers and engineers has created a global collaboration, called CORD-19, which collects thousands of scientific publications focused on the new coronavirus \cite{Wang2020CORD19TC}.

\begin{figure*}[h]
\begin{subfigure}{.5\textwidth}
	\centering
	\includegraphics[width=7cm]{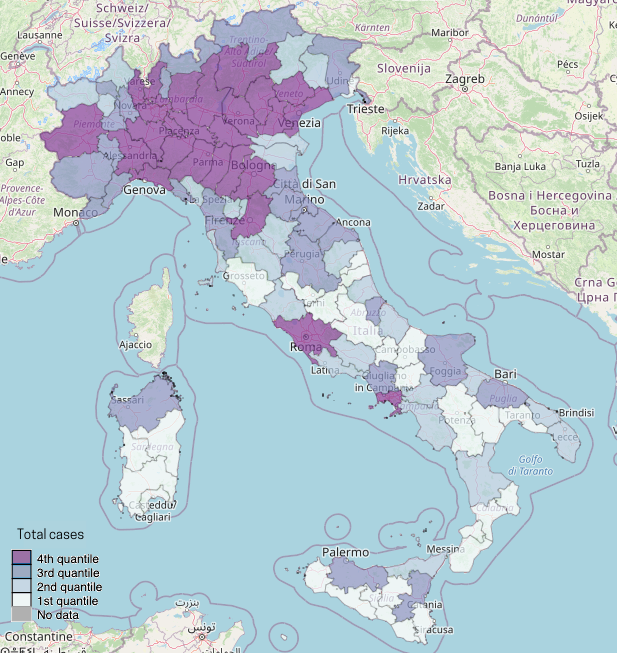}
	\caption{}
	\label{fig:total_cases}
\end{subfigure}
\begin{subfigure}{.49\textwidth}
	\centering
	\includegraphics[width=7cm]{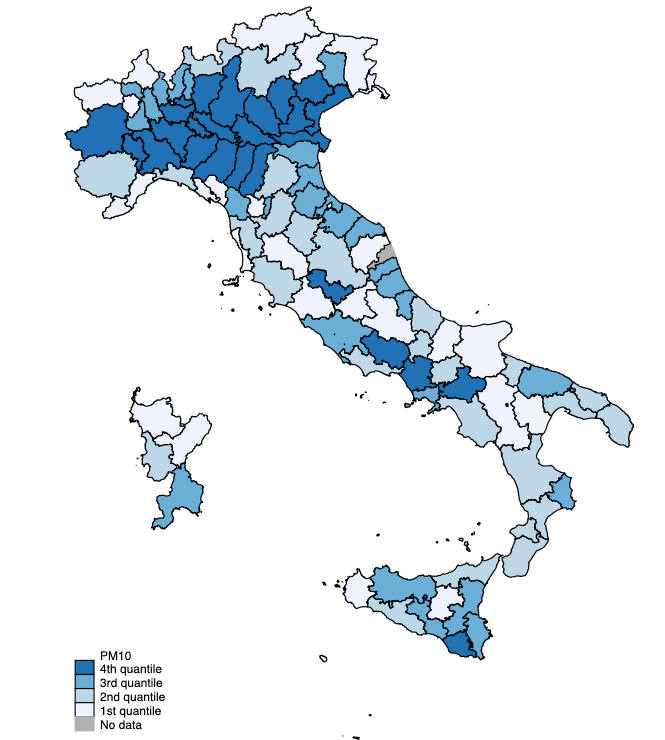}
	\caption{}
	\label{fig:PM10_italia}
\end{subfigure}
\caption{(\ref{fig:total_cases}) The choropleth map shows the distribution of COVID-19 total cases in the Italian districts (as of $3^{rd}$ October 2020). The data are divided into quartiles. (\ref{fig:PM10_italia}) The choropleth map shows the average PM$_{10}$ concentrations of 2018 in the Italian districts. The data are divided into quartiles. \cite{becchetti2020understanding}}
\end{figure*}

Identifying the main factors that contribute to the spread of the SARS-CoV-2 virus is certainly one of the current public health goals. However, the complexity of the phenomenon and the limited availability of information make this study particularly difficult. The possible relationship between fine particulate matter (PM) and SARS-CoV-2 immediately aroused particular interest, especially if one compares the distribution of infections and pollutants. 

\begin{figure}
	\centering
	\includegraphics[width=\linewidth]{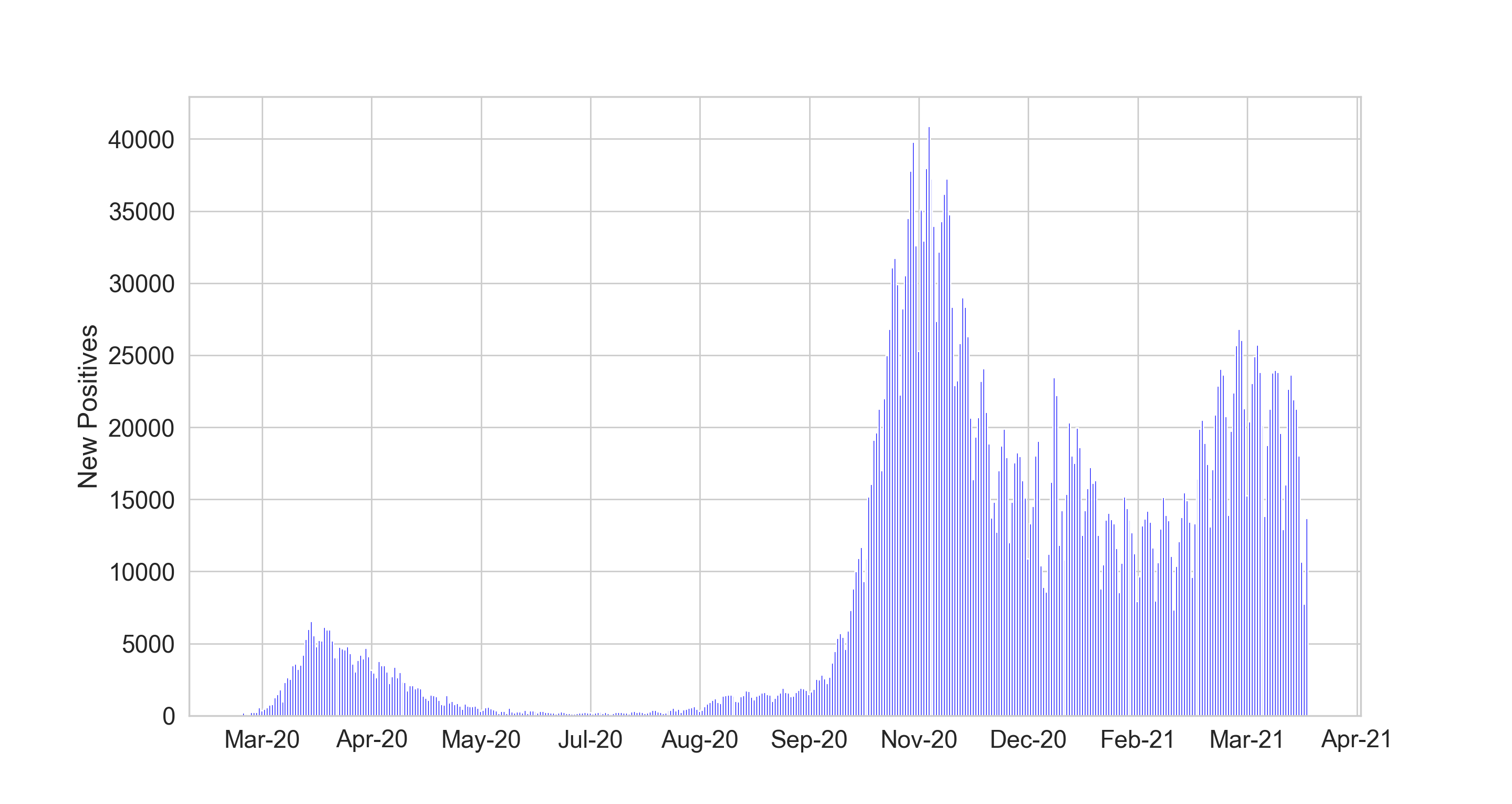}
	\caption{Number of diagnosed cases of COVID-19 in Italy by test/diagnosis date (from 24th of February 2020 to 8th of April 2021).}
	\label{fig:casi_per_prelievo}
\end{figure}

In \cite{becchetti2020understanding}, authors show that areas with high concentration of pollutants (i.e., dark blue districts in Figure \ref{fig:PM10_italia}) mostly coincide with areas with high number of positive cases of COVID-19 (i.e., Figure \ref{fig:total_cases}).
Unfortunately, the time window used in the two figures is different, mostly due to a lack of information for Italian districts in 2019 and 2020 about the pollutants. However, this observation motivates the interest in investigating the possible correlations between air quality and daily confirmed cases of COVID-19. In addition, the graph of the new daily cases in Italy (Figure \ref{fig:casi_per_prelievo}) shows an important decrease in infections during the summer months. This leads us to think that there may be a relationship between the environmental temperature and the spread of the pandemic.
In this work we focus on the study of possible relationships among the number of new daily infected cases and the air quality in some Italian districts. 

\textbf{Contribution. } In this work, 
we performed a cross-correlation analysis that highlights possible relationships among the number of daily cases and several factors related to the air quality. 
We exploit these correlations by developing and comparing three different supervised learning models. We trained these models to predict the number of new cases of COVID-19, showing that the number of infected cases can be computed in advance with good accuracy. These tools can be used to enrich the set of information available to governments and institutions for helping in making decisions in order to protect the population and stemming the pandemic. Indeed, accurate predictive models might help modeling possible scenarios, helping government institutions to better manage the pandemic.

\section{Related Works}
The rapid spread of the COVID-19 pandemic has attracted the attention of numerous scholars and researchers from many different disciplines. The aim is twofold: on one side, scholars want to understand the modalities of transmission of the SARS-CoV-2 virus and its mechanisms of interaction with the host. On the other side, they want to investigate all possible contributing causes that may have played a key role in the number of infections and in the mortality rate of the disease. Currently, evidence indicates that the SARS-CoV-2 virus spreads mainly from person to person through the inhalation of respiratory droplets, which are normally released when an infected person speaks, coughs or sneezes \cite{la2020coronavirus}. However, it is hypothesized that the virus may be aerosolized during certain activities or procedures and may remain active for prolonged periods \cite{van2020aerosol}.
On the $30^{th}$ of January 2020, the Istituto Superiore di Sanità (ISS) confirmed the first two cases of SARS-CoV-2 infection in Italy: two tourists from Wuhan landed in Milan and then hospitalized in Rome. The first autochthonous positive case was confirmed on the $21^{st}$ of February 2020 and was a patient hospitalized in serious condition in Lodi. Always on the $21^{st}$ of February 2020 the first death of COVID-19 in the country was reported; he was a man from Vo' (Padova). From the $23^{rd}$ of February 2020, 11 municipalities in northern Italy (in Lombardy and Veneto) were quarantined and from the $10^{th}$ of March 2020 the lockdown was extended to the entire country, until  the $3^{rd}$ of June 2020. Of course, the taken measures may have influenced the progress of the pandemic. The study by Lavezzo et al. \cite{lavezzo2020suppression}, for example, shows that containment measures have helped to decrease the transmission of SARS-CoV-2 in the municipality of Vo'.

The review of the literature, conducted on the variables that influence seasonal viruses affecting the respiratory system, underlines that there is a multitude of factors that can influence the behaviour of viruses, including humidity, pollution, human behaviour, physiological and demographic characteristics, human mobility, as well as climate change \cite{sloan2011impact}. Each of these factors is important as it affects virus survival, virulence, and transmissibility between individuals. From several studies, that have examined the mechanisms underlying the seasonal nature of respiratory viral infections, it has been deduced that, in general, the two main factors contributing to the spread of virus infections are identifiable in changes in environmental parameters and in human behaviours. A recent investigation \cite{moriyama2020seasonality,lolli2020impact}, that has  analysed the mechanisms of action of viruses, shows how the combination of favourable winter levels of humidity, temperature, and solar radiation can compromise our antiviral defense mechanisms, resulting in a greater susceptibility of the host to respiratory viruses. Furthermore, various studies \cite{ciencewicki2007air,xing2016impact,grigg2018air} report evidence in favour of an association between exposure to air pollutants and the increased risk of respiratory viral infections, although the potential cellular and molecular mechanisms underlying the increased susceptibility are still largely unknown.

The scientific literature that has investigated the possible relationships between environmental factors and SARS-CoV-2 is very large and there are also conflicting opinions. In \cite{setti2020evaluation} Setti et al. show how the PM$_{10}$ limit exceedances may be compatible with a role of particulate matter as virus carrier. This hypothesis is also supported by the discovery of the presence of SARS-CoV-2 RNA on atmospheric particulate matter \cite{setti2020sars}. Furthermore, the results of a survey on 120 Chinese cities \cite{yongjian2020association} reveal significantly positive associations between daily measurements of atmospheric particulate matter and nitrogen dioxide and COVID-19 confirmed cases, while sulphur dioxide is negatively associated.

Regarding ozone, many studies \cite{murray2008virion,hudson2009development,tseng2008inactivation} claim that it is particularly lethal against viruses due to its high oxidizing property. However, there are no studies confirming the role of ozone in the specific inactivation of SARS-CoV-2. Anyway, it was effective in killing the SARS-CoV virus of the 2003 epidemic \cite{zhang2004examination} and therefore it could also be lethal against SARS-CoV-2 as both viruses come from the same group and have similar structures. This hypothesis, however, would not seem to agree with the results of a recent research \cite{yongjian2020association} that found a significantly positive association between ozone concentrations and daily COVID-19 confirmed cases.

Finally, the scientific literature argues that high temperature and relative humidity affect the environmental resistance of SARS-CoV-2, reducing its spread. Two different studies \cite{magurano2020sars,chin2020stability} show that virus viability decayed more rapidly at higher temperatures, indicating that viral infectivity can be altered with the increase of temperature. Furthermore, a recent research \cite{sobral2020association} claims the existence of a negative correlation between the average temperature by country and the number of SARS-CoV-2 infections. Regarding relative humidity, another study \cite{wang2020high} supports the existence of robust negative associations between humidity and transmission of COVID-19.

\section{Background}
In this work, we performed an analysis on the collected data, studying the correlation between environmental features and the target variable (i.e., the number of new daily infected cases). 

\subsection{Correlation Analysis}
A correlation analysis \cite{FRANZESE2019706} is a statistical study that evaluates the strength and the sign of a relationship between two variables. There exist many different indexes that can be employed to describe such relationships.
The Pearson's correlation coefficient \cite{benesty2009pearson} $ r_p $ is usually adopted as correlation index. Given two vectors of values $ \mathbf{X} $ and $ \mathbf{Y} $, the Pearson correlation index can be calculated as follows:

\begin{displaymath}
	r_p = \frac{\displaystyle \sum_{i=0}^{n-1}(x_{i}-\hat{x})(y_{i}-\hat{y})}{\displaystyle \sqrt{ \sum_{i=0}^{n-1}(x_{i}-\hat{x})^{2}(y_{i}-\hat{y})^{2}}}
\end{displaymath}

where $ n $ is the number of samples, $x_{i}$ and $y_{i}$ are the samples,  and $ \hat{x} $ and $ \hat{y} $ correspond to the average values of $\mathbf{X}$ and $\mathbf{Y}$ respectively.

This index assumes that: 
(i) both variables are normally distributed;
(ii) relationship between each of the two variables is linear;
(iii) variables have continuous values;
(iv) data are equally distributed about the regression line (also known as homoscedasticity). 

Another function is the so-called Spearman correlation coefficient \cite{myers2004s}. This is a simple and efficient way to analyze the similarity of the shape of two time series, it is computed as follows:

\begin{displaymath}
	r = 1 - \frac{\displaystyle 6 \sum d_{i}^2}{n(n^2-1)}
\end{displaymath}

where 
$d_{i}$ is the difference between the two ranks of each observation and $n$ is the number of observations. Spearman coefficient operates on raw data, it is based on the ranks of the data, it is insensitive to outliers and can operate with ordinal values. 
Due to the characteristics of the data, in this work, we adopted the Spearman correlation coefficient.

The value of the coefficient is in $[-1,1]$ and it describes the relationship between the variables. Specifically, a high value, close to $+1$, indicates a positive correlation, while a low value, close to $-1$, indicates a negative correlation. A positive correlation exists when the increase in the value of one variable makes also increase the value of the other variable. On the other hand, if a negative correlation exists, then as the value of one variable increases, the value of the other variable decreases.
Furthermore, when the index value is close or equal to $0$ there is a poor or no correlation between the two variables, which means that increasing or decreasing one variable does not affect the value of the other variable.
In particular, we will use the following terminology based on the absolute value of $r$, we say that there is 
no correlation or very weak correlation when $r < 0.3$; weak correlation when $0.3 \leq r <0.5$; moderate when $0.5 \leq r < 0.7$; and we will say that the correlation is strong when $r \geq 0.7$.

It is important to remember that the correlation analysis does not provide any indication of a cause-effect relationship between the variables. To establish a true causal condition, the variables should be completely isolated from any other possible confounding variable. If a correlation is found between air quality and SARS-CoV-2 infection, this would constitute just one more proof to be able to subsequently support any scientific demonstration.

\subsection{Machine Learning Techniques}
Machine learning is a branch of AI that studies and develops learning algorithms able to model intrinsic characteristics or relationships in the data. 
Usually, machine learning algorithms have a bottom-up approach, which means that they infer information from a collection of data called dataset, which describes the studied scenario. Thus, a dataset is an M $ \times $ N matrix in which each column corresponds to a variable (also called “feature”) that describes a specific characteristic of the domain, and each row corresponds to a sample $c_{i}= (x_{i};y_{i})$ where $x_{i}= (x_{i, 1}, x_{i, 2},\ldots, x_{i, N } )$ and $y_{i} $ represents the “label” of the sample (i.e., the value of a variable that we want to predict), which is not always known a priori. The problem addressed in this paper is a supervised learning problem. The term “supervised” refers to the fact that in the set of samples the labels $y_{1}, \ldots, y_{M}$ are already known. In this approach, we assume that there exists an ideal function $ f: X \rightarrow Y $ such that $ f(x_{i}) = y_{i} $, where $ X $ is the space of all possible samples and $ Y $ is the set of all possible outputs. Supervised learning tries to find a function $ \bar {f}: X \rightarrow Y $ that approximates $ f $ as closely as possible, finding the same labels as $ f $ for most of the samples.

In this work we adopt the following models:
Random Forest, XGBoost, and Neural Network. 

\subsubsection{Random Forest}
The Random Forest algorithm \cite{breiman2001random} is based on decision trees and applies the bagging technique. This consists in training multiple decision trees on distinct partition of the dataset by sub-sampling it with re-insertion. 
Random Forest typically has better generalization performance than a single decision tree, thanks to randomness which helps to contain the over-fitting problem, reducing model variance.

\subsubsection{XGBoost}
XGBoost is based on Friedman's original Gradient Boosting \cite{friedman2002stochastic,friedman2001greedy}. It introduces a regularization term to control over-fitting, obtaining better performance results. The Gradient Boosting technique creates a final model based on a combination of single models such as Random Forest, but it builds them sequentially by giving more weight to instances with incorrect predictions. Specifically, in each learning cycle, prediction errors are used to calculate the gradient, i.e. the partial derivative of the loss function with respect to the prediction, and build a new tree capable of predicting gradients. Then, the prediction values are updated. After the learning phase, XGBoost derives the final predictions of the target variable by adding the average calculated in the initial step to all the residuals predicted by the trees, multiplied by the learning rate.

\subsubsection{Neural Network}
Neural Network \cite {shalev2014understanding} is made up of stacks of neurons. A neuron is a processing unit connected to different inputs $ x_{i} $, each of which is associated with a weight $ w_{i} $. The neuron computes the weighted sum of the input vector $ z = w_{1} x_{1} + w_{2} x_{2} + ... + w_{m} x_{m} = \mathbf {w^{ T} x} $ and applies a threshold function $ \sigma $ to it. The set of nodes of the network can be decomposed into a union of disjoint subsets $ V_{0}, V_{1} ..., V_{T} $, called “layers”, such that each edge connects some node in $ V_{t-1} $ to some node in $ V_{t} $. The lower level, $ V_{0} $, is called “input layer”, the levels $ V_{1}, ..., V_{T-1} $ are called “hidden layers”, while the upper level $ V_{T} $ is called “output layer”. During the learning phase, the neural network adopts a back-propagation approach, which repeatedly updates the weights, minimizing the loss function. At the end of this phase, it uses the forward propagation to calculate the prediction.

\section{Empirical Study}
In this section, we describe the datasets and the results of the correlation analysis performed on them. We describe  the predictive models and their performances on the different datasets.

\subsection{Data Collection}
The data used in this work comes from two different sources: one contains the daily details of the pandemic, and the other contains the environmental information of different districts or geographical areas around the world.

We focus on the pandemic situation in Italy. Italian data about the pandemic has been made available in a GitHub repository under a CC-BY-4.0 license from the Italian Civil Protection Department (ICP) \footnote{https://github.com/pcm-dpc/COVID-19 - Last visited on 8th of April 2021}. In this repository, the number of total cases is available at the level of each Italian district. For security and privacy reasons, other information (i.e., the number of infected cases in a specific city or the number of deaths per district) is stored  and protected in a platform of the Integrated Surveillance and thus accessible only by authorized people \footnote{https://www.epicentro.iss.it/en/coronavirus/sars-cov-2-integrated-surveillance-data - Last visited on 8th of April 2021}. Therefore, only the daily number of total cases of COVID-19 in different areas is used to describe the progress of the pandemic in Italy. 

The time window of this study goes from the $1^{st}$ of January 2020 
to the $8^{th}$ of April 2021 (date of our last measurement). 
We collected environmental data from the Air Quality Open Data Platform (AQODP) \cite{world2020air}. This platform was created by the World Air Quality Index project team and contains meteorological and air quality information of major cities around the world, unfortunately not all the data published in this website is validated. But it is worth noting that data of Italian districts contained in AQODP is provided by the ARPA (Agenzia Regionale per la Protezione Ambientale), which is an official and trusted source for this kind of data. 
The AQODP platform publishes information about 12 districts of Italy. Due to the fact that many features of 4 out of the 12 districts have missing values, we decided to focus exclusively on the eight districts with the most complete set of data. Specifically, we used data about Bologna, Brescia, Milan, Modena, Naples, Parma, Prato, and Rome. 
For each of these districts, we derived a dataset merging data from the two aforementioned sources: each row in a dataset describes information about environmental factors and the number of new infected cases for a specific date. All the adopted variables are summarized in Table \ref{tab:legend}.

\begin{table*}[h]
	\begin{center}
		\caption{List of considered variables in the datasets.}
	\label{tab:legend}
		\begin{tabular}{r p{5cm} cc}
			\itshape{Variable} & \itshape{Description} & \itshape{Measure} & \itshape{Source} \\
			\hline
			date & Date &&\\
			humidity\_median & Daily median of the relative humidity & percentage & ARPA \\
			no2\_median & Daily median concentration of NO$_{2}$ (nitrogen dioxide) & $ \mu g / m^{3} $ &  ARPA \\
			o3\_median & Daily median concentration of $O_{3}$ (ozone) & $ \mu g / m^{3} $ &  ARPA \\
			pm10\_median & Daily median concentration of PM$_{10}$ & $ \mu g / m^{3} $ &  ARPA \\
			pm2.5\_median & Daily median  concentration of PM$_{2.5}$ & $ \mu g / m^{3} $ &  ARPA \\
			so2\_median & Daily median concentration of SO$_{2}$ (sulfur dioxide) & $ \mu g / m^{3} $ &  ARPA \\
			temp\_median & Daily median temperature & Celsius &  ARPA \\
			total\_cases & Cumulative number of COVID-19 cases & &  ICP \\
			new\_cases & Number of new daily cases of COVID-19& &  ICP \\
			\hline
		\end{tabular}
	\end{center}
\end{table*}

\subsection{Data Analysis}


Initially, we performed a pre-processing task which removed all negative values for the $new\_cases$ variable. These negative values occur when the Italian Civil Protection Department adjusted the daily data about total cases for some areas, resulting in a reduced  number of infected people compared to the number of the previous day. This was probably due to errors in the positive cases count. In addition, all records with missing values were removed before the correlation analysis.

In order to perform an accurate correlation analysis, it is necessary to take into account a probable incubation period of the virus. It is important to notice that an additional delay time might be due to the bureaucracy related to the execution and analysis of the nasopharyngeal swab. This value was not known to us a priori. 

In this study, cross-correlations are used for the analysis of time-lagged relationships between several environmental factors and their possible influence on the number of new positive cases. The use of the cross-correlation functions allows to assess the sensitivity and responsiveness at different time \cite{derrick2004time}. This is due to the fact that a specific environmental configuration may influence the spread of the virus, but consequences may be evident only some days later.  For a specific factor evidence, the amount of days needed is not known a priori.
Therefore, to find the best time-lag, we shift the number of new daily cases of $i$ positions in the datasets (with $i$ varying from $0$ to $60$). This is done to compare the environmental data of a given day with the number of new infected cases after $i$ days. In this way, we looked for the maximum correlation value of each environmental parameter in a time window of two months in the past.

\subsection{Correlation Analysis Results}
In general, we observed a strong negative correlation with temperature and ozone; a moderate positive correlation with $NO_{2}$, $PM_{2.5}$,  $PM_{10}$, and humidity; a poor positive correlation with $SO_{2}$. Depending on the area, the results are different.
Among all the analyzed districts, here we report results for Brescia dataset and a brief discussion about the differences with other datasets. All the results about the correlation analysis are available under the Appendix in Section \ref{app:corr_results}.

\begin{figure*}[htp]
	\centering
	\includegraphics[width=1\linewidth]{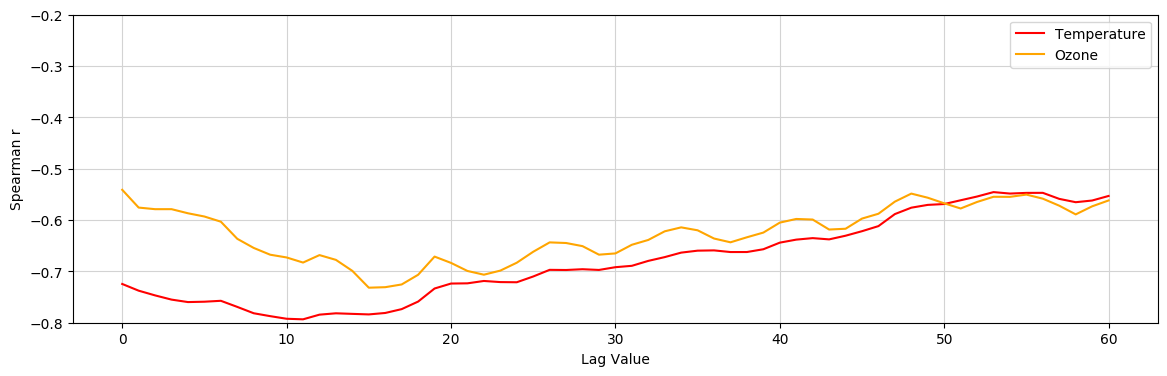}
	\caption{Area of Brescia: cross-correlations values for temperature, ozone, and new daily infected cases, sliding time-window from $0$ to $60$.}
	\label{fig:corr_brescia_in}
\end{figure*}

In the district of Brescia (which was one of the most compromised during the pandemic), we found a strong negative correlation with the temperature and a negative correlation with ozone. 
This means that as the daily temperature increases, we observe that the number of daily infections decreases. In particular, the correlation peak between temperature and new COVID-19 cases occurred for $i = 10$ ($ r = -0.7799$, $ p\_value < 0.001$).
Similarly, ozone has a strong negative correlation with COVID-19 cases and the pick is for $i = 15$ ($ r = -0.7358$, $p\_value < 0.001$). Therefore, an increase in the maximum concentration of ozone in the atmosphere is associated with a decrease in positive cases.
Figure \ref{fig:corr_brescia_in} depicts the different values for the correlation indexes when we shift the time-lag window from $0$ to $60$ days. As you can notice, after the peak the effect of these two factors on the virus decreases as expected.

\begin{figure*}[htp]
	\centering
	\includegraphics[width=1\linewidth]{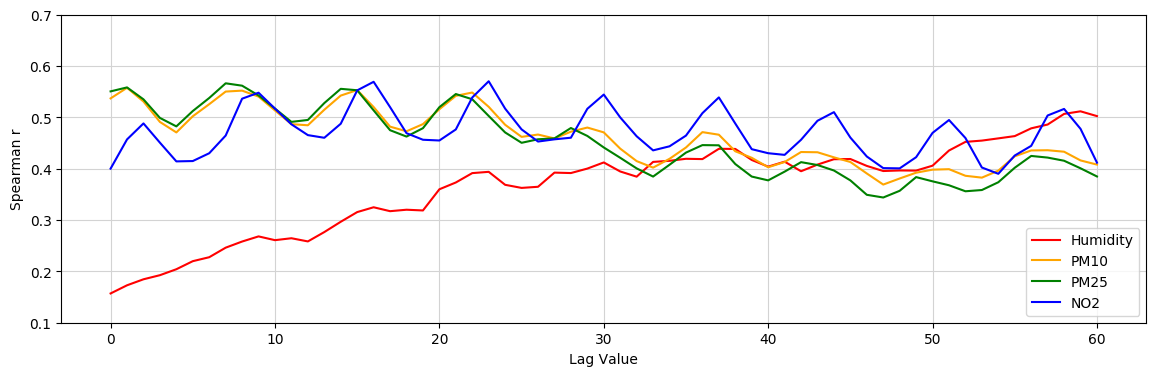}
	\caption{Area of Brescia: cross-correlations values for $NO_{2}$, $PM_{2.5}$, $PM_{10}$, and new daily infected cases, sliding time-window in $0$ to $60$.}
	\label{fig:corr_pm_brescia_in}
\end{figure*}

On the other hand, the number of new infected cases and the concentration of the main air pollutants (i.e., $NO_{2}$, $PM_{2.5}$, and $PM_{10}$) show a moderate correlation at close but different lag values. Specifically, the median concentrations of $NO_{2}$, $PM_{2.5}$ and $PM_{10}$ are positively correlated and the values of the correlation coefficients are respectively: for $NO_{2}$ after $i =23$ days, $r = 0.5699$ ($p\_value < 0.001$); for $PM_{2.5}$ after $i =21$ days, $r = 0.5639$ ($p\_value < 0.001$); and for $PM_{10}$ after $i=22$ days, $r = 0.5617$ ($p\_value < 0.001$). These results indicate that a higher median daily concentration of these pollutants is associated with a greater number of people contracting the infection after more or less 20 days. Figure \ref{fig:corr_pm_brescia_in} depicts the different values for correlation indexes when we shift the time-lag window from $0$ to $60$ days. The oscillating behaviour that can be observed in Figure \ref{fig:corr_pm_brescia_in} might be due to traffic emissions or industrial productions which are higher during working days of the week and should decrease in the week-end (also due to the restrictions imposed). In fact, oscillations have a period of 7 days.
Similar results were also obtained with data regarding the areas of Milan, Bologna, Parma and Modena. 

The results obtained for the datasets of Naples, Prato, and Rome are much weaker (see Section \ref{app:corr_results} in the Appendix). The moderate correlation with temperature and ozone might be caused by a set of co-factors probably not considered in this study. For instance, all the other considered areas are in the Po Valley which is a geographical area surrounded by the Alps and the Apennines. The wind is rare and the air is colder in the plains than in the mountains, causing emissions stay above Po Valley and making harder for natural and artificial emissions to be dissolved.  
This may be one of the reasons why the atmospheric pollutants seem to have more effects on the spread of the virus in areas of Po Valley than in other Italian areas which instead are close to the sea like Rome or Naples.
Although the sign of the correlation coefficients are in line with those of the other northern provinces. 



\begin{figure*}[htbp]
	\centering
	\includegraphics[width=\linewidth]{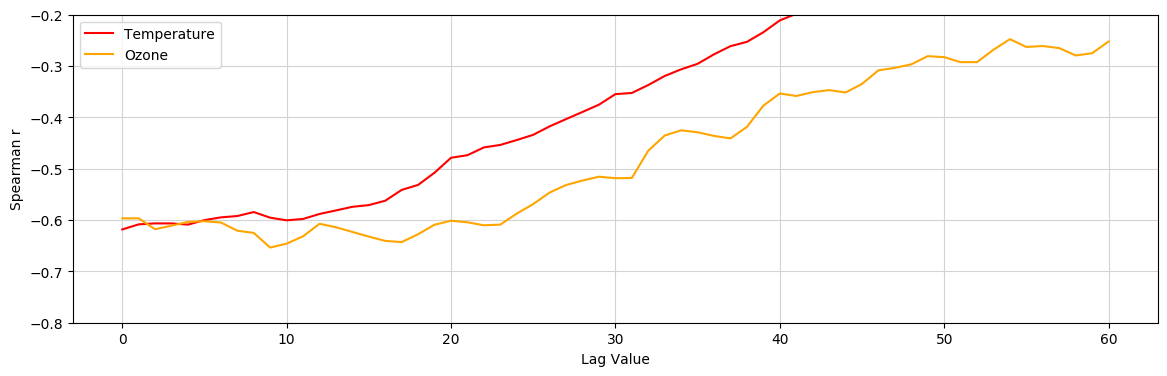}
	\caption{Area of Rome: cross-correlations values for temperature, ozone, and new daily infected cases, sliding time-window from $0$ to $60$.}
	\label{fig:corr_roma_in}
\end{figure*}
\begin{figure*}[htbp]
	\centering
	\includegraphics[width=\linewidth]{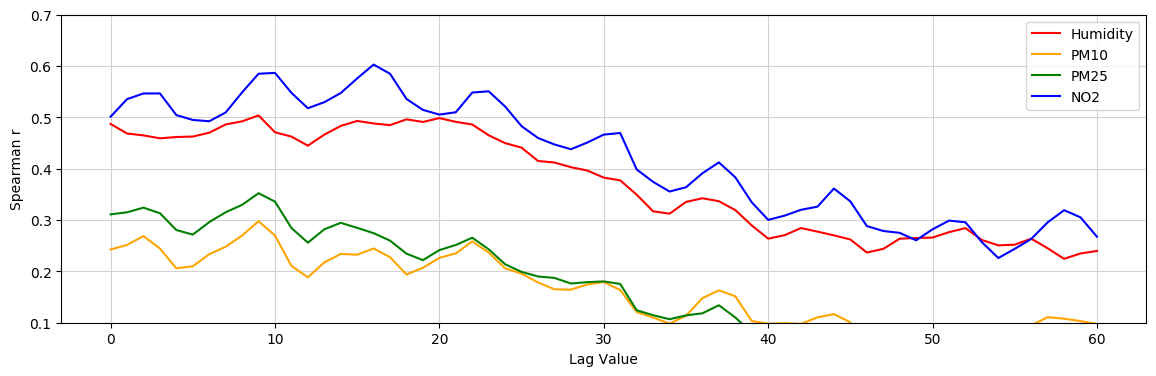}
	\caption{Area of Rome: cross-correlations values for temperature, ozone, and new daily infected cases, sliding time-window from $0$ to $60$.}
	\label{fig:corr_pm_roma_in}
\end{figure*}

The correlation peak occurs at distance of different days depending on the parameter analyzed: atmospheric particulate matter, median nitrogen dioxide and ozone are more related to new cases registered after 22-23 days from the environmental measurements, while temperature is more related to cases identified 9-10 days later. This difference could be due to a different effect of these factors on the virus. Indeed, atmospheric pollutants could take a few days to decrease the environmental resistance of SARS-CoV-2, while the contribution of the temperature to the spread of the virus could be more immediate. 

\subsection{Machine Learning Models}
Given the statistically significant correlation of some variables with COVID-19 cases, we developed and trained some regressors for predicting the number of new infected cases based on the values of the environmental parameters. The machine learning algorithms used in this work estimate the impact of the environmental factors on the spread of the COVID-19 pandemic. The aim of these models is to predict the number of confirmed cases given the measurements of the atmospheric variables.

The simulations are developed in Python 3.7. We adopted \textit{RandomForestRegressor} by Scikit-learn, \textit{XGBRegressor} by XGBoost and \textit{Sequential Model} by Keras for Neural Network.

\begin{table}[htp]
	\begin{center}
	\caption{Neural network configuration.}
	\label{tab:nnconfig}
		\begin{tabular}{lllc}
			Layers & Shape & Learnable &  \itshape{Act. Function}\\
			\hline
Dense &             (None, 100)     &          1300   &ReLu \\   
Dense &             (None, 50)      &          5050   &ReLu \\   
Dense &            (None, 10)    &            510   &ReLu \\       
Dense &             (None, 1) &                11    & Linear\\    
\hline
\multicolumn{2}{l}{Total params} & \multicolumn{2}{l}{6,871}\\
\multicolumn{2}{l}{Trainable params} & \multicolumn{2}{l}{6,871}\\
\hline
		\end{tabular}
	\end{center}
\end{table}

The configuration of the used neural network is reported in Table \ref{tab:nnconfig}. Moreover, for this model data were standardized using \textit{StandardScaler()} by Scikit-learn. 
Each model was trained separately for each of the eight districts for both the value of $i$ corresponding to the pollutant correlation peak (e.g., $i=22$) and for that corresponding to the temperature correlation peak (e.g., $i=10$). This was done to compare the results from different trainings in order to determine which is the best delay time for predicting new cases in each district.

In order to estimate the generalization performance, we adopted the holdout approach, i.e. the dataset was divided into two disjoint sets, called training sets and test sets. The training set contains 70\% of the instances of the original dataset and it was used to train the model. The test dataset contains the remaining 30\% of the samples, and it was used to test the generalization level of the regressor. The split of the instances was done randomly.

For \textit{RandomForestRegressor} a tuning phase of the hyperparameters was performed in order to obtain the best possible accuracy. To do that, we adopted a grid search approach 
(i.e., a list of allowed values is specified for each hyperparameters and then they are evaluated through a 5-fold cross validation to determine the best combination). 
We chose to optimize the following parameters: 

\begin{itemize} 
    \item \textit{max\_depth}, that is the maximum depth that each tree can have. Values for this parameter were searched in the interval [3,7)
    \item \textit{n\_estimators}, that is the number of trees. Values for this parameter were searched in $\{10, 50, 100, 1000\}$.
\end{itemize}

In order to evaluate and compare the different models, we compute the Root Mean Squared Error (RMSE), the Mean Absolute Error (MAE), and the R$^{2}$ score on the test set. 
Assuming that $\hat{y_{i}}$ is the predicted value of the i-th sample and $y_{i}$ is the corresponding observed value, then the three metrics are defined as follows:
\begin{align*}
	& RMSE(y,\hat{y}) = \sqrt{\frac{1}{n}\sum_{i=0}^{n-1} (y_{i}-\hat{y_{i}})^2}
	\\
	& MAE(y,\hat{y}) = \frac{1}{n}\sum_{i=0}^{n-1} |y_{i}-\hat{y_{i}}|
	\\
	& R^2(y,\hat{y}) = 1- \frac{\sum_{i=1}^{n} (y_{i}-\hat{y_{i}})^2}{\sum_{i=1}^{n} (y_{i}-\bar{y})^2}
\end{align*}

where $ n $ is the number of samples and $ \bar{y}$ is the average of the observed values of the target variable. The quadratic exponent in the RMSE allows to heavily penalize  large errors. However, this makes Root Mean Squared Error more sensitive to outliers than Mean Absolute Error.

RMSE and MAE are measures of error, therefore when we compare two regression models on the same dataset, the one with the lowest values is the one with the best predictions. In contrast, $ R^2 $ score or coefficient of determination represents the proportion of variance of $y$ that has been described by the independent variables of the model. This metric provides an indication of the goodness of fit and thus it is a measure of the likelihood that samples never seen by the model are predicted correctly. The best possible score is 1 and it occurs when it is possible to predict exactly what the value of the target variable will be, knowing the values of the independent variables. A constant model that always predicts the expected value of $y$, ignoring the input features, has a $ R^2 $ score equal to 0. The value of $ R^2 $ can also be negative as the model can be arbitrarily worse than a constant model. Therefore, if we compare two regression models on the same dataset, the model with the greater $ R^2 $ score will be the one with the highest predictive power.

Furthermore, the Mean 5-Fold Cross Validation score was computed over the entire dataset. This latter approach of evaluation combines the 5-fold cross validation technique with the $ R^2 $ score in order to obtain a more generalized result.

\subsection{Prediction Results}
\begin{figure}[htbp]
	\centering
	\includegraphics[width=\linewidth]{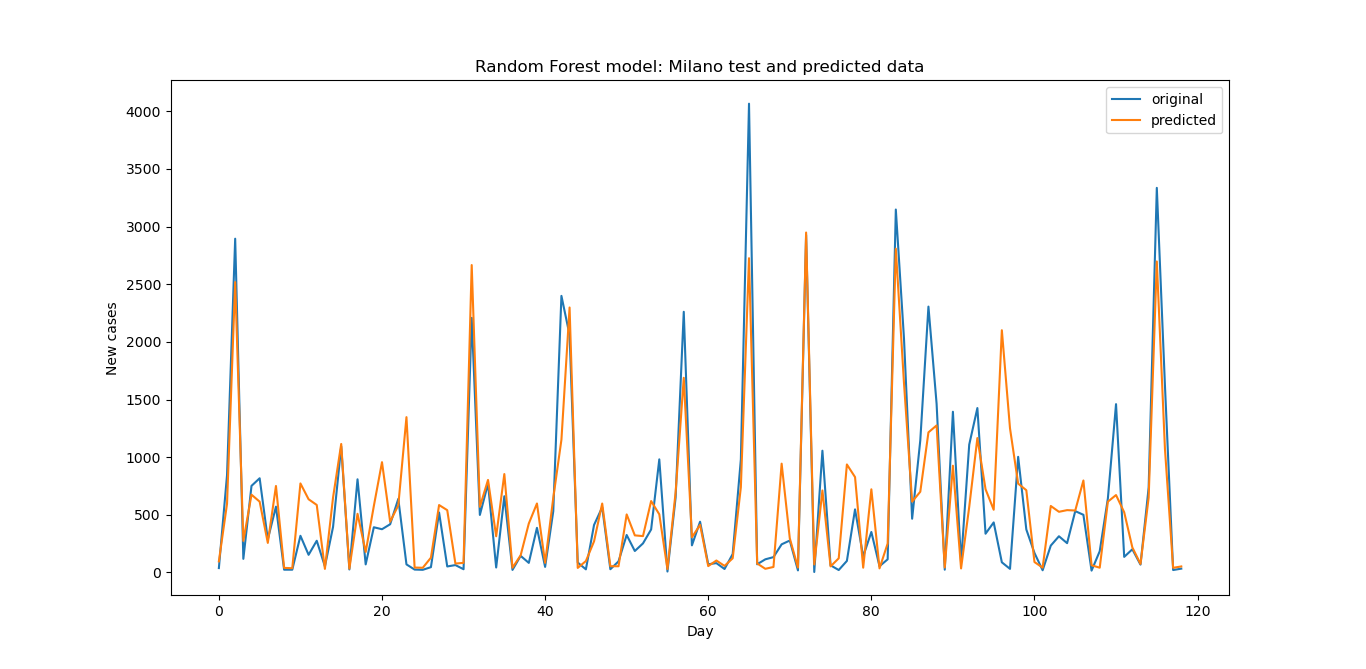}
	\caption{RandomForest on the area of Milan: predicted number of new infected cases compared with the real number of infected cases. $ i = 10 $.}
	\label{fig:milan10ggrf}
\end{figure}

As expected, the results obtained for each province are different. The models have greater predictive capacity for the area of Milan (Table 3), followed by Bologna. The maximum performance using Milan dataset was achieved with the Random Forest algorithm. The best delay time due to the identification of the disease was found to be 10 days (although good results were obtained even with i = 15). Figure \ref{fig:milan10ggrf} shows the comparison graph of the daily cases of the province of Milan observed and predicted by Random Forest (when $ i = 10 $ and the Cross Validation accuracy is 73 \%, $ DS = 2 \% $). Observing the figure, it can be seen that the predictions are very good, even if in some days the predicted cases are slightly higher than those observed.

\begin{table*}[htbp]
	\begin{center}
	\caption{Performance of the models on the dataset of the province of Milan.}
		\begin{tabular}{clcccc}
			$i$ value & \textit{Model} & \textit{R$ ^{2} $ score} & \textit{RMSE} & \textit{MAE} & \textit{5-fold CV score} \\
			\hline
			10 & XGBoost & 0,55 & 530,08 & 280,62 & 0,66 $ \pm $ 0,10 \\
			10 & Neural Network & 0,67 & 454,51 & 277,60 & 0,67 $ \pm $ 0,10 \\
			\textbf{10} & \textbf{Random Forest} & \textbf{0,74} & \textbf{405,36} & \textbf{241,27} & \textbf{0,73 $ \pm $ 0,02} \\
			15 & XGBoost & 0,71 & 447,70 & 265,24 & 0,65 $ \pm $ 0,13 \\
			15 & Neural Network & 0,75 & 441,84 & 262,80 & 0,65 $ \pm $ 0,10 \\
			15 & Random Forest & 0,66 & 518,65 & 283,73 & 0,72 $ \pm $ 0,07 \\
			21 & XGBoost & 0,36 & 719,91 & 364,29 & 0,54 $ \pm $ 0,20 \\
			21 & Neural Network & 0,41 & 684,29 & 374,53 & 0,49 $ \pm $ 0,16\\
			21 & Random Forest & 0,46 & 660,96 & 338,90 & 0,53 $ \pm $ 0,18 \\
			\hline
		\end{tabular}
	\end{center}
	\label{tab:risultati_milano}
\end{table*}

Instead, for Bologna district the maximum Cross Validation accuracy was 70\% ($ DS = 5 \% $) by adopting XGBoost as predictive model and 16 days as delay time.
On the other hand, performances of Brescia, Modena, Naples, Parma, Prato and Rome are weaker, with Cross Validation accuracy values ranging between 36 \% and 63 \%. In general, the models that perform better are XGBoost and Random Forest, while the neural network performs worst. 

The low accuracy values of the models could hide a dependence of the target variable, that is the number of daily COVID-19 cases in Italy, on many factors, probably not observable from the available data. Indeed, it must be taken into account that the main modality of transmission of the virus is direct contact between people. Consequently, the behavior of the population and containment measures could affect the number of infections. Another important factor is certainly the number of swabs carried out daily in each district. Unfortunately this information is not available at district level.

\section{Limitations}
In this paper we tried to understand, on the basis of current knowledge and  available data, whether air quality may play a role in the spread of the COVID-19 pandemic and, in particular, on the number of recorded daily cases in several Italian districts. Of course, the spread of a viral infection is a complex and multi-factor system. Therefore, our analysis includes some limitations:
\begin{itemize}
    \item \textbf{Open-source datasets}: open source datasets containing official information at district level are very little, poor or completely missing for some areas. Although numerous research studies on the COVID-19 pandemic have been published so far, in most cases the databases used have not been made available or they include data only at national level, such as the well-known dataset by Johns Hopkins University \footnote{https://coronavirus.jhu.edu/map.html - Last visited on 18th of January 2021 }. Furthermore, almost all the information at provincial level about the spread of the pandemic in Italy is stored in a platform accessible only to the Istituto Superiore della Sanità (ISS) and other authorized entities.

    \item \textbf{Data accuracy}: the Civil Protection Department has repeatedly corrected past data, published in the repository, modifying the daily data. Of course, the inaccuracy in the number of documented infected can cause a significant increase in the uncertainty of the estimate provided by the prediction models based on historical data. In addition, environmental measurements may also be subject to error. Indeed, the World Air Quality Index project team has underlined that not all data have been validated \cite{world2020air}.
    
    \item \textbf{Missing dimensions}: the employed machine learning models does not take into account (or at least very marginally) the time dimension. We only consider the data for a given day and we are not taking into account any restriction that the government enacted during the pandemic. We presume that the information of previous days may influence the daily data. Thus considering the time dimension might improve the performance of regressors.
    
    \item \textbf{Confounding factors}:
factors capable of generating spurious associations, which could have altered the results. For instance, the restrictive measures and the rigor with which they have been observed is a possible confounding factor. As well as the number of daily swabs which varied considerably over the time.

    \item \textbf{Lack of knowledge}:  this work is based on a still uncertain understanding of the phenomenon. There are many questions that the research has yet to answer. For example, it is not clear to what extent surfaces and aerosols favor the transmission of the virus and whether it is actually possible that SARS-CoV-2 can travel incorporated into air pollution particles while maintaining its vitality.

    \item \textbf{Data availability, quality and representativeness}: machine learning algorithms require a large amount of data to be able to accurately learn the relationships between variables. A small dataset could be poorly representative of the variability of interactions and could consequently lead to low predictive performance. The limited observation period, due to the recent discovery of the virus, could therefore represent a further limitation of the empirical study.
\end{itemize}

\section{Conclusion}

In this work, we investigate possible relationships among environmental parameters, geographical distribution and the spread of the COVID-19 pandemic in different Italian areas. The analysis highlights a possible diagnostic delay period. It has also shown that machine learning techniques can be applied to make useful predictions on the number of COVID-19 cases per day as a function of environmental data measurements. For instance, the possibility of predicting future new infected cases could be useful to make adequate decisions on the management of the pandemic, avoiding the overload of the health system. 

This work can be seen as another step towards the understanding of a complex system, which deserves to be investigated through in-depth scientific studies. Future epidemiological investigations should be based on sufficiently extensive and comprehensive data. In addition, further studies aimed at investigating the possible mechanisms of interaction of environmental factors with SARS-CoV-2 are needed.

In the future, the analysis will be extended to other geographical areas and additional co-factors will be included in the dataset in order to improve the performance of the models. We plan to investigate the application of new ensemble methods (e.g., \cite{Cornelio2021}) to improve performance, and recurrent neural networks to take into account the time dimension in the analysis.



\bibliography{biblio}

\newpage
\appendix

\section{Correlation Results}
\label{app:corr_results}
This appendix reports the results of the correlation analysis for each dataset. For the sake of readability, we restrict the plot to the median value of each attribute. The plot limits are deliberately kept fixed for an easiest comparison among them.

 \subsection{Bologna dataset}
\begin{figure*}[htp] 	\centering 	\includegraphics[width=1\linewidth]{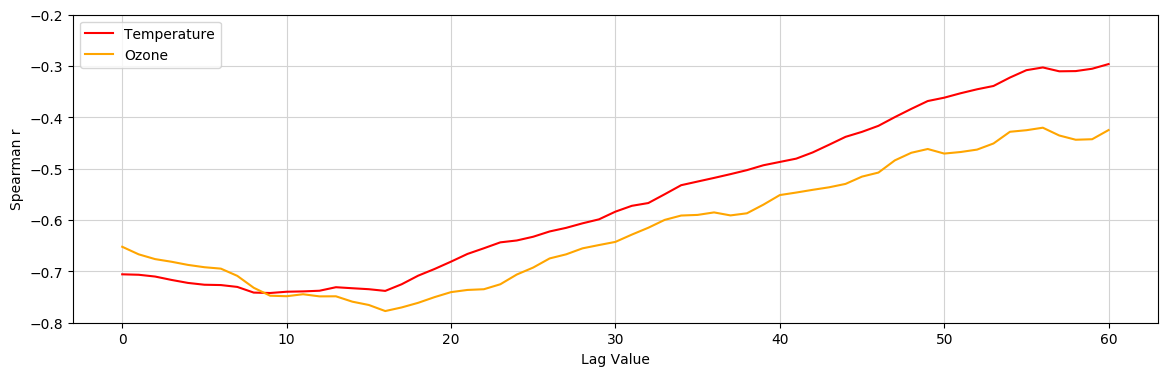} 	\caption{Area of Bologna: cross-correlations values for temperature 
, ozone, 
 and new daily infected cases, sliding time-window from $0$ to $60$.} 	\label{fig:corr_bologna} \end{figure*}

\begin{figure*}[htp]	\centering	\includegraphics[width=1\linewidth]{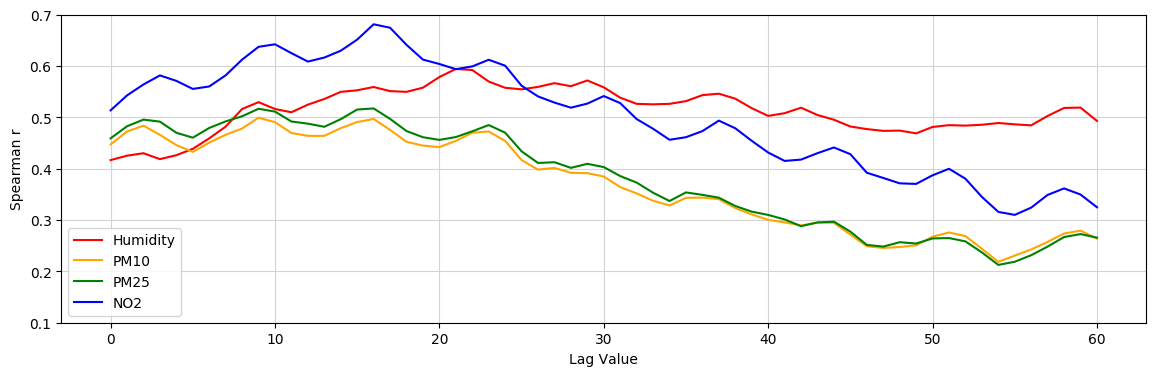}	\caption{Area of Bologna: cross-correlations values for $NO_{2}$, $PM_{2.5}$, $PM_{10}$, and new daily infected cases, sliding time-window in $0$ to $60$.}	\label{fig:corr_pm_bologna} \end{figure*}

This section reports the results of the correlation analysis for the data set of Bologna. 		Figure \ref{fig:corr_bologna} shows the correlation for the maximal values of temperature 
 and ozone.
	Figure \ref{fig:corr_pm_bologna} shows the correlation for the median values of humidity and $NO_{2}$, and for the 		maximal values for $PM_{2.5}$ and $PM_{10}$.
\newpage \subsection{Brescia dataset}
\begin{figure*}[htp] 	\centering 	\includegraphics[width=1\linewidth]{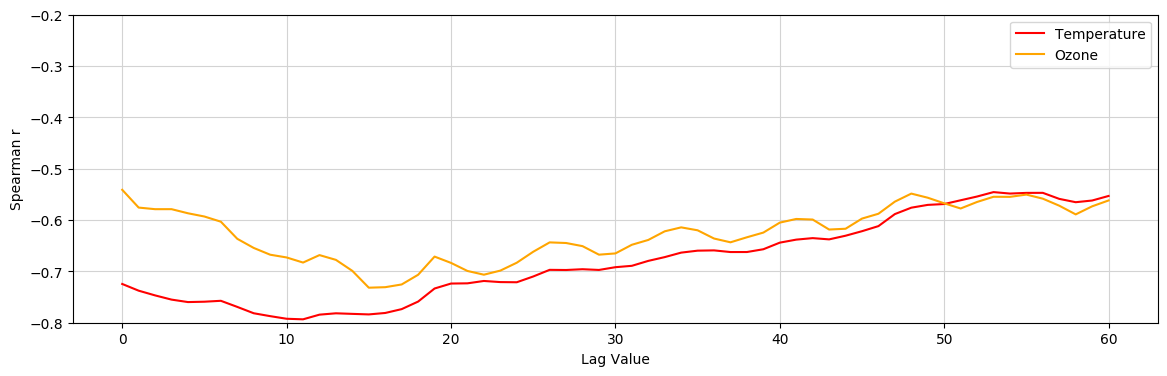} 	\caption{Area of Brescia: cross-correlations values for temperature 
, ozone, 
 and new daily infected cases, sliding time-window from $0$ to $60$.} 	\label{fig:corr_brescia} \end{figure*}

\begin{figure*}[htp]	\centering	\includegraphics[width=1\linewidth]{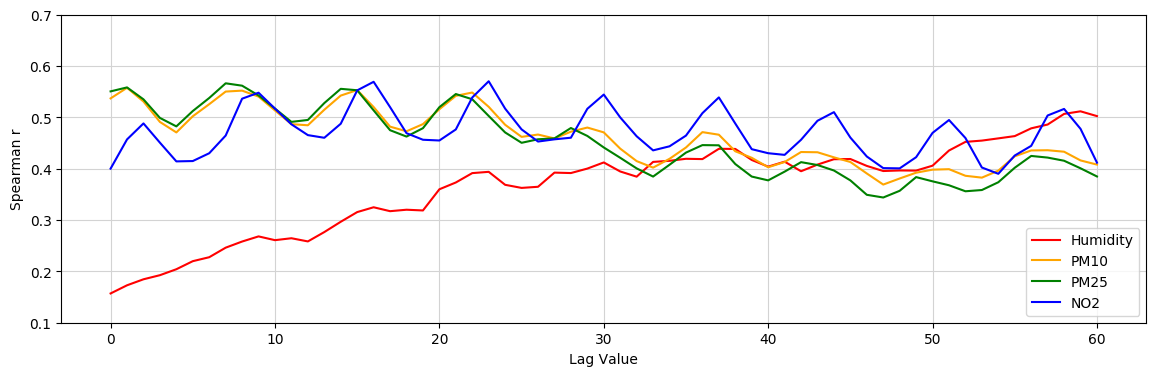}	\caption{Area of Brescia: cross-correlations values for $NO_{2}$, $PM_{2.5}$, $PM_{10}$, and new daily infected cases, sliding time-window in $0$ to $60$.}	\label{fig:corr_pm_brescia} \end{figure*}

This section reports the results of the correlation analysis for the data set of Brescia. 		Figure \ref{fig:corr_brescia} shows the correlation for the maximal values of temperature 
 and ozone.
	Figure \ref{fig:corr_pm_brescia} shows the correlation for the median values of humidity and $NO_{2}$, and for the 		maximal values for $PM_{2.5}$ and $PM_{10}$.

\newpage \subsection{Milan dataset}
\begin{figure*}[htp] 	\centering 	\includegraphics[width=1\linewidth]{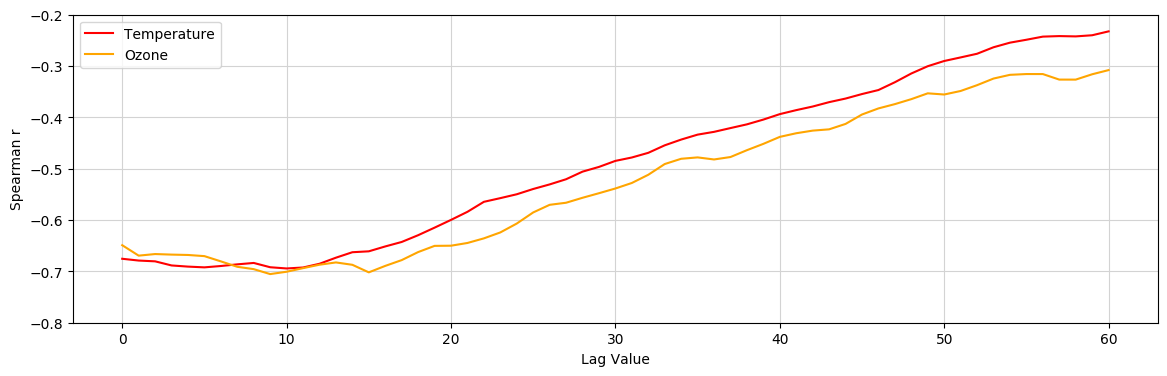} 	\caption{Area of Milan: cross-correlations values for temperature 
, ozone,
 and new daily infected cases, sliding time-window from $0$ to $60$.} 	\label{fig:corr_milan} \end{figure*}

\begin{figure*}[htp]	\centering	\includegraphics[width=1\linewidth]{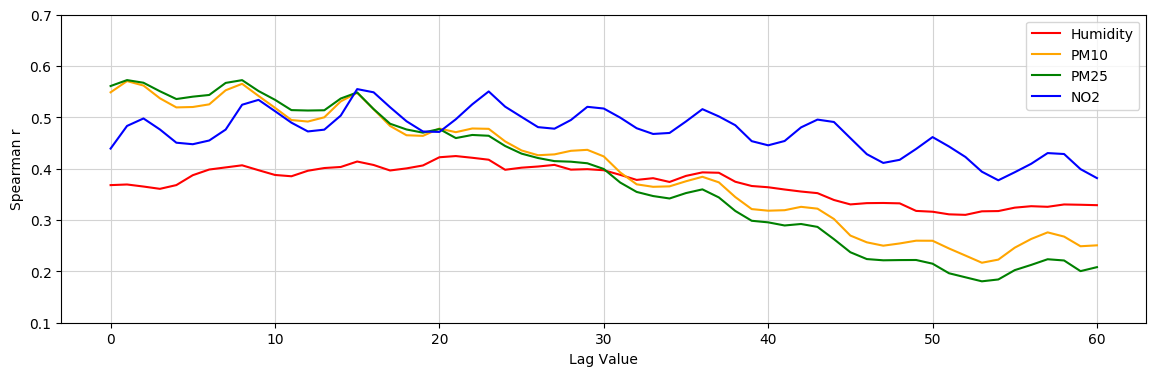}	\caption{Area of Milan: cross-correlations values for $NO_{2}$, $PM_{2.5}$, $PM_{10}$, and new daily infected cases, sliding time-window in $0$ to $60$.}	\label{fig:corr_pm_milan} \end{figure*}

This section reports the results of the correlation analysis for the data set of Milan. 		Figure \ref{fig:corr_milan} shows the correlation for the maximal values of temperature 
 and ozone.
	Figure \ref{fig:corr_pm_milan} shows the correlation for the median values of humidity and $NO_{2}$, and for the 		maximal values for $PM_{2.5}$ and $PM_{10}$.
\newpage \subsection{Modena dataset}
\begin{figure*}[htp] 	\centering 	\includegraphics[width=1\linewidth]{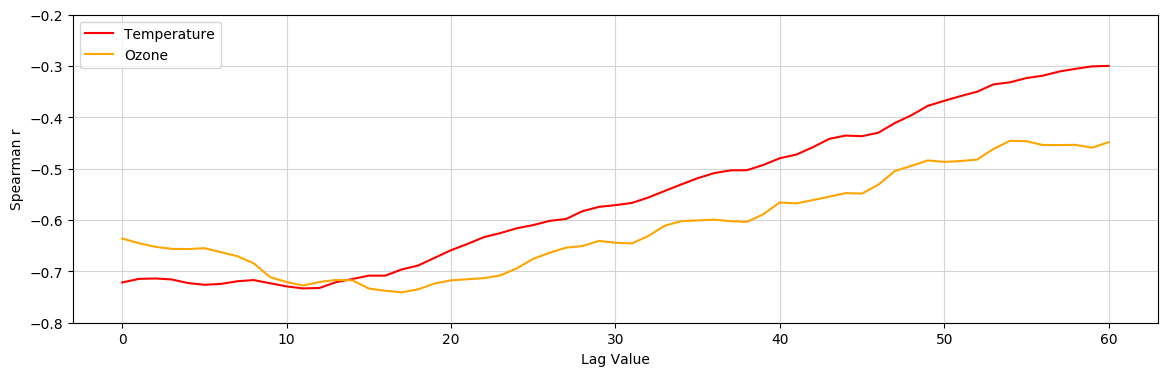} 	\caption{Area of Modena: cross-correlations values for temperature 
, ozone, 
 and new daily infected cases, sliding time-window from $0$ to $60$.} 	\label{fig:corr_modena} \end{figure*}

\begin{figure*}[htp]	\centering	\includegraphics[width=1\linewidth]{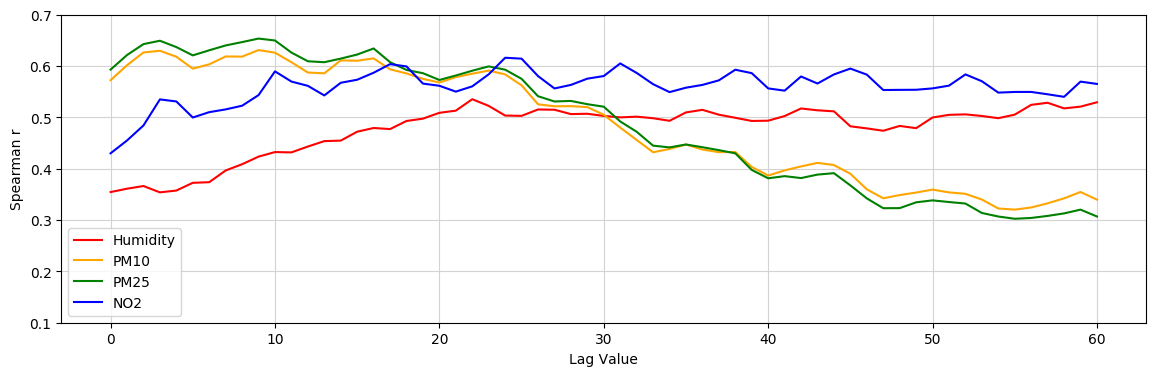}	\caption{Area of Modena: cross-correlations values for $NO_{2}$, $PM_{2.5}$, $PM_{10}$, and new daily infected cases, sliding time-window in $0$ to $60$.}	\label{fig:corr_pm_modena} \end{figure*}

This section reports the results of the correlation analysis for the data set of Modena. 		Figure \ref{fig:corr_modena} shows the correlation for the maximal values of temperature 
 and ozone.
	Figure \ref{fig:corr_pm_modena} shows the correlation for the median values of humidity and $NO_{2}$, and for the 		maximal values for $PM_{2.5}$ and $PM_{10}$.
\newpage \subsection{Naples dataset}
\begin{figure*}[htp] 	\centering 	\includegraphics[width=1\linewidth]{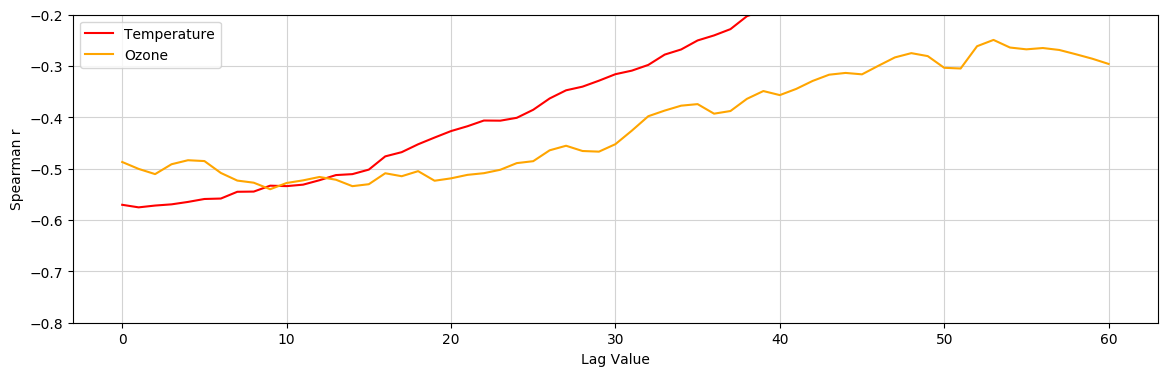} 	\caption{Area of Naples: cross-correlations values for temperature 
, ozone, 
 and new daily infected cases, sliding time-window from $0$ to $60$.} 	\label{fig:corr_naples} \end{figure*}

\begin{figure*}[htp]	\centering	\includegraphics[width=1\linewidth]{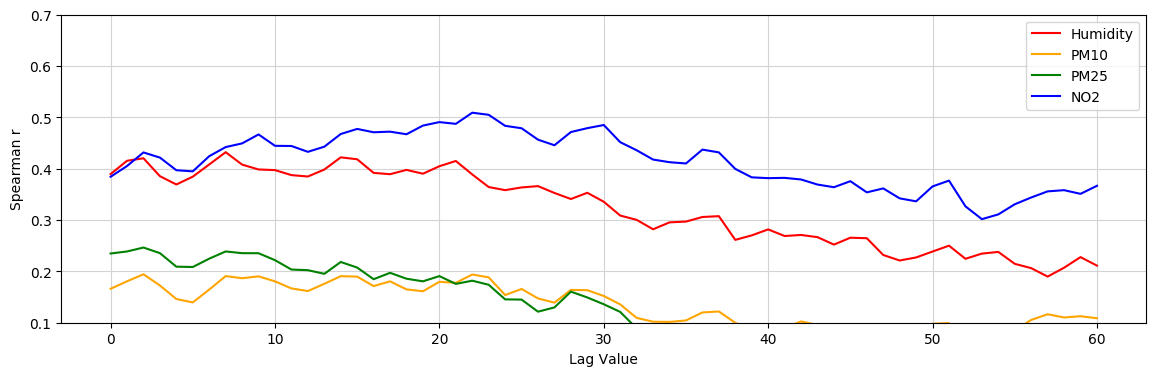}	\caption{Area of Naples: cross-correlations values for $NO_{2}$, $PM_{2.5}$, $PM_{10}$, and new daily infected cases, sliding time-window in $0$ to $60$.}	\label{fig:corr_pm_naples} \end{figure*}

This section reports the results of the correlation analysis for the data set of Naples. 		Figure \ref{fig:corr_naples} shows the correlation for the maximal values of temperature 
 and ozone.
	Figure \ref{fig:corr_pm_naples} shows the correlation for the median values of humidity and $NO_{2}$, and for the 		maximal values for $PM_{2.5}$ and $PM_{10}$.
\newpage \subsection{Parma dataset}
\begin{figure*}[htp] 	\centering 	\includegraphics[width=1\linewidth]{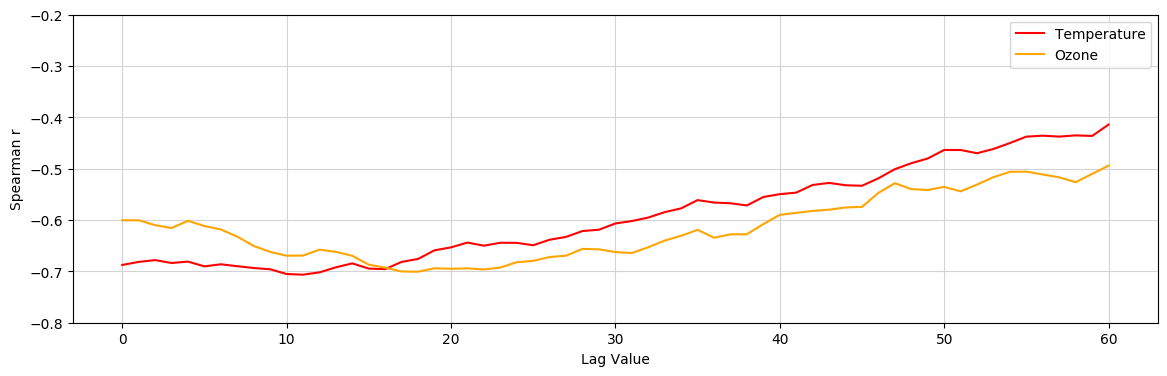} 	\caption{Area of Parma: cross-correlations values for temperature 
, ozone, 
 and new daily infected cases, sliding time-window from $0$ to $60$.} 	\label{fig:corr_parma} \end{figure*}

\begin{figure*}[htp]	\centering	\includegraphics[width=1\linewidth]{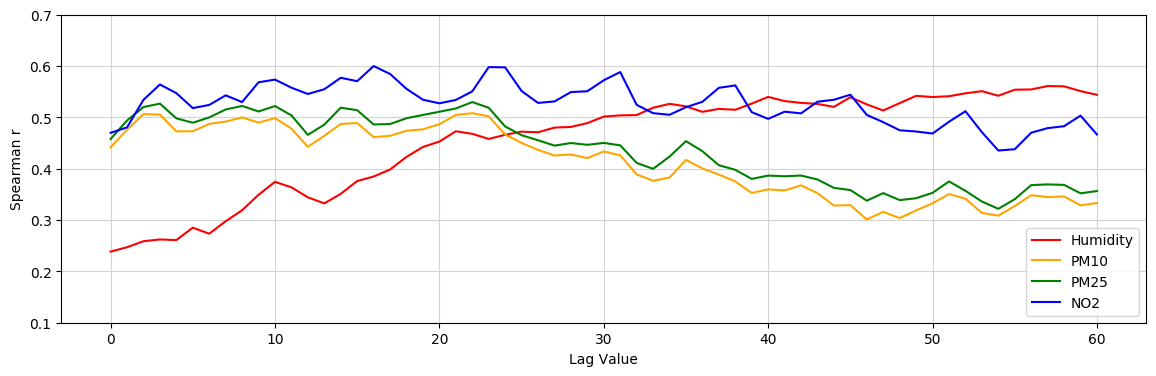}	\caption{Area of Parma: cross-correlations values for $NO_{2}$, $PM_{2.5}$, $PM_{10}$, and new daily infected cases, sliding time-window in $0$ to $60$.}	\label{fig:corr_pm_parma} \end{figure*}

This section reports the results of the correlation analysis for the data set of Parma. 		Figure \ref{fig:corr_parma} shows the correlation for the maximal values of temperature 
 and ozone.
	Figure \ref{fig:corr_pm_parma} shows the correlation for the median values of humidity and $NO_{2}$, and for the 		maximal values for $PM_{2.5}$ and $PM_{10}$.
\newpage \subsection{Prato dataset}
\begin{figure*}[htp] 	\centering 	\includegraphics[width=1\linewidth]{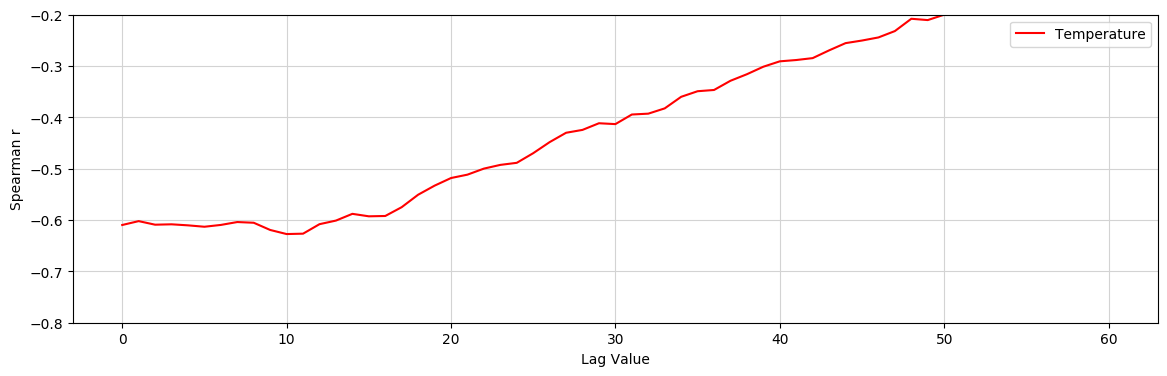} 	\caption{Area of Prato: cross-correlations values for temperature 
 and new daily infected cases, sliding time-window from $0$ to $60$.} 	\label{fig:corr_prato} \end{figure*}

\begin{figure*}[htp]	\centering	\includegraphics[width=1\linewidth]{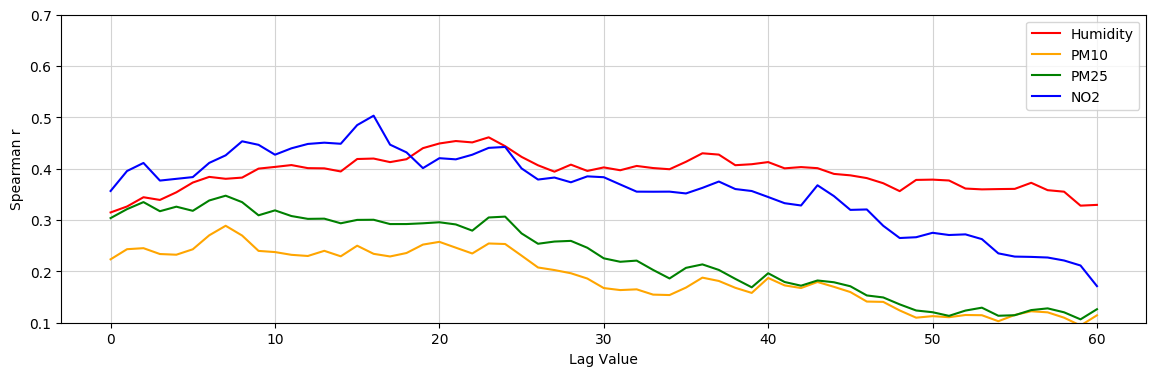}	\caption{Area of Prato: cross-correlations values for $NO_{2}$, $PM_{2.5}$, $PM_{10}$, and new daily infected cases, sliding time-window in $0$ to $60$.}	\label{fig:corr_pm_prato} \end{figure*}

This section reports the results of the correlation analysis for the data set of Prato. 		Figure \ref{fig:corr_prato} shows the correlation for the maximal values of temperature 
 (data for ozone not available for this dataset).
	Figure \ref{fig:corr_pm_prato} shows the correlation for the median values of humidity and $NO_{2}$, and for the 		maximal values for $PM_{2.5}$ and $PM_{10}$.
\newpage \subsection{Rome dataset}
\begin{figure*}[htp] 	\centering 	\includegraphics[width=1\linewidth]{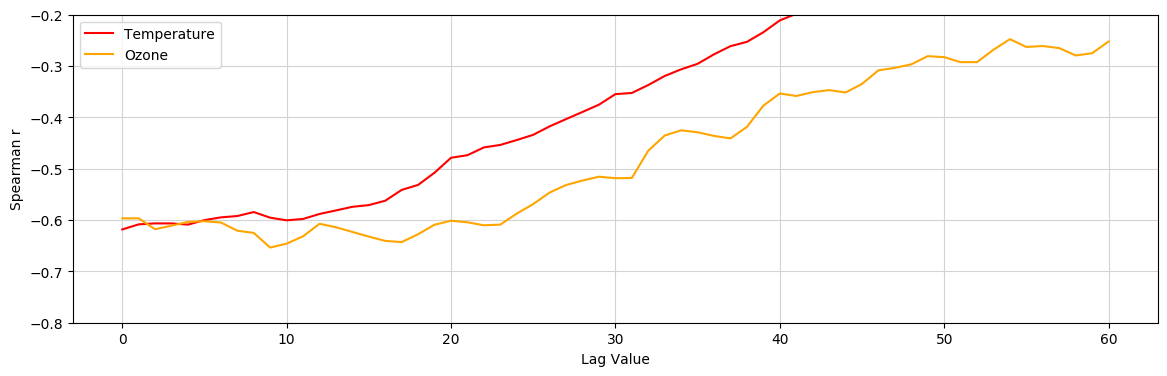} 	\caption{Area of Rome: cross-correlations values for temperature 
, ozone,
 and new daily infected cases, sliding time-window from $0$ to $60$.} 	\label{fig:corr_rome} \end{figure*}

\begin{figure*}[htp]	\centering	\includegraphics[width=1\linewidth]{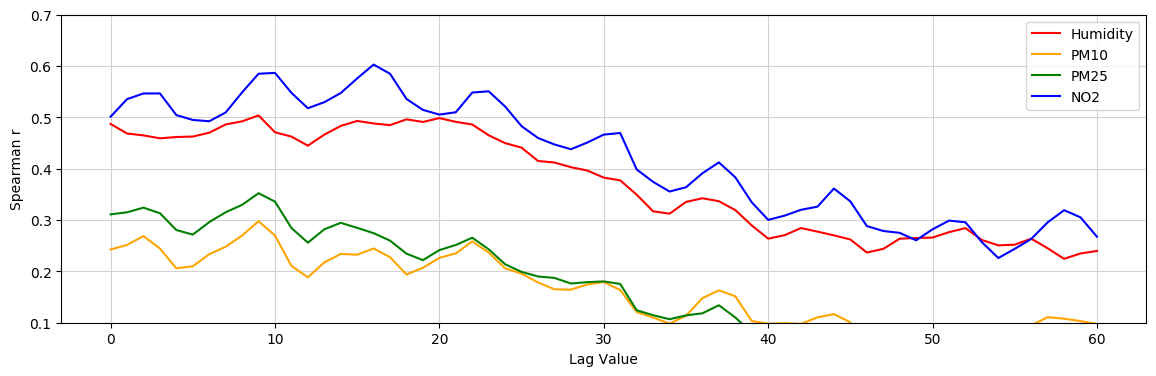}	\caption{Area of Rome: cross-correlations values for $NO_{2}$, $PM_{2.5}$, $PM_{10}$, and new daily infected cases, sliding time-window in $0$ to $60$.}	\label{fig:corr_pm_rome} \end{figure*}

This section reports the results of the correlation analysis for the data set of Rome. 		Figure \ref{fig:corr_rome} shows the correlation for the maximal values of temperature 
 and ozone.
	Figure \ref{fig:corr_pm_rome} shows the correlation for the median values of humidity and $NO_{2}$, and for the 		maximal values for $PM_{2.5}$ and $PM_{10}$.

\end{document}